\documentclass[english]{achemso}
\usepackage[T1]{fontenc}
\usepackage[latin9]{inputenc}
\usepackage{amstext}
\usepackage{graphicx}

\makeatletter

\title{On the elasticity of phantom model networks with cyclic defects}
\author{Michael Lang}
\affiliation{Leibniz Institute of Polymer Research Dresden, Hohe Straße 6, 01069
Dresden, Germany}
\email{lang@ipfdd.de}

\makeatother

\usepackage{babel}
\begin{document}
\begin{abstract}
The impact of finite cycles on the phantom modulus in an otherwise
perfect network is computed exactly. It is shown that pending cycles
reduce the phantom modulus of the network by $kT/V$ independent of
junction functionality. The correction for non-pending cycles is larger
than estimated previously within this particular approximation of
the surrounding network structure. It is discussed that loop formation
inevitably leads to streched chain conformations, if the loops are
built step by step as part of the network structure. All network loops
tend to contract simultaneously to optimize conformations, which leads
to an increasing stretch of chains in larger loops that can be observed
in computer simulations. Possible other corrections to phantom modulus
that were left aside in previous work are discussed briefly.
\end{abstract}
Understanding the elasticity of a polymer network or gel is one of
the key problems of polymer physics. Nowadays, it is generally accepted
that rubber elasticity stems from two major contributions: the entanglement
modulus, $G_{\text{e}}$, caused by the non-cross-ability of the polymer
strands and the phantom modulus, $G_{\text{ph}}$, that resembles
the contribution of network connectivity at the absence of entanglements.
The exact solution of the phantom model of a perfect defect free model
network containing no finite cyclic structures (``loops'') is discussed
in textbooks on polymer physics, see for instance, ref. \cite{Rubinstein2003}.
It has been shown that linear and branched network defects can be
treated by estimating the cycle rank \cite{Flory1982} of the network.
But in contrast to the assumptions made in the textbook derivation,
real polymer networks are made entirely of finite cycles, with a small
average number of chains per cycle \cite{Lang2001,Lang2007}. Therefore,
the generalization of our understanding of the phantom modulus from
a perfect tree to a more realistic structure containing finite cycles
is an open problem of polymer physics since the last 40 years \cite{Flory1976}.

Recently, first steps were made towards a solution of this problem
\cite{Zhong2016} by considering the impact of isolated cyclic structures
within an otherwise perfect tree (``ideal loop gas approximation'',
ILGA) made of $f$-functional junctions and ideal strands of $N$
segments between. Even though the ILGA is not realistic concerning
the particular structure of the surrounding network, it should allow
at least for a qualitative trend of the impact of finite cycles on
elasticity. Quite interestingly, the results of ref. \cite{Zhong2016}
were in clear disagreement to the result discussed in previous work
\cite{Flory1982} or textbooks \cite{Rubinstein2003}. It was found
that finite loops lead generally to a reduction of modulus. In particular,
modulus was reduced by even more than $kT$ per pending cycle (cyclic
structure with only a single connection to the rest of the network)
for small junction functionality $f\le4$. If these results were correct,
it would question whether the cycle rank is the correct concept to
treat network defects at least for $f\le4$. And it would call for
an improved estimate of the phantom modulus as proposed in ref. \cite{Zhong2016}.

In the present paper, several problems related to the phantom modulus
of a network containing finite cycles are addressed. Detailed computations
are provided in the text document of the supporting information, while
the main text provides an enhanced discussion of the results. The
computation starts with the ILGA of ref \cite{Zhong2016}, which is
solved here exactly for loops of arbitrary size. These results are
compared with an alternate derivation considering explicitly the sequence
of attaching strands to form a loop. It is shown that the process
of network formation leads to stretched chain conformations once finite
loops are formed inside the network by a step-wise addition of strands.
These stretched conformations can be identified in simulation data
providing evidence that cross-link fluctuations and contribution to
modulus decouple for phantom networks made of finite cycles. Other
mechanisms that might impact the phantom modulus are stressed briefly
at the end of the manuscript.

The basic idea of the computations in section S1-S4 of the supporting
informations is that the fluctuations of any part of the network around
its average position in space can be modeled by the end-motion of
a virtual chain that is attached to the non-fluctuating elastic background
(EB) on its other end \cite{Rubinstein2002}. The particular structure
of the phantom network enters by introducing correlations among the
fluctuations of connected molecules and for computing the size of
the virtual chain that connects to EB. 

\begin{figure}
\includegraphics[width=0.9\columnwidth]{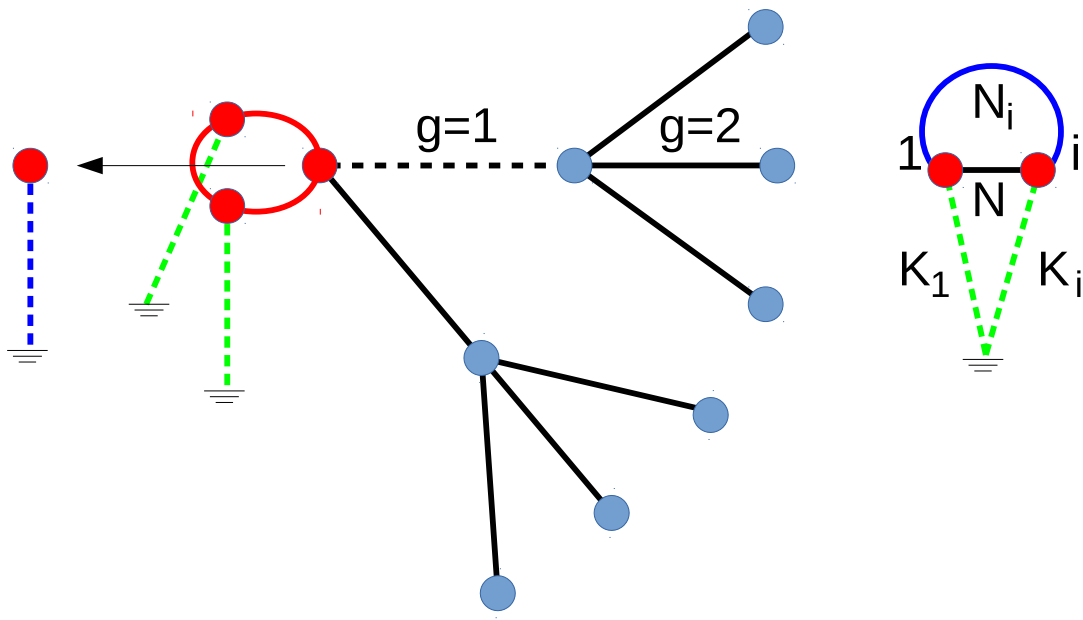}

\caption{\label{fig:Illustration-of-the}Left part: The fluctuations of each
cross-link (red dots) inside the red loop are modeled by a virtual
chain (blue dashed line) that connects to EB (ground symbol). The
green dashed lines indicates the connections through the network to
the EB outside of the loop. The first two generations of attached
strands, $g=1$ and $g=2$, and the corresponding junctions on the
way to EB (of the infinite tree to EB) are shown explicitly for one
cross-link by black lines and gray dots. Right: reduced loop structure.
One $N$-mer that connects junction $1$ and $i$ directly, the strand
$N_{\text{i}}$ through the remainder of the loop but not through
EB plus two strands $K_{1}$ and $K_{i}$ that connect junction 1
and $i$ through EB.}
\end{figure}

We are interested in the elastic effectiveness $\epsilon$ of a strand
in the network, which is defined here as the contribution of this
strand to the shear modulus of the network in units of $kT/V$. Here
$k$ is the Boltzmann constant, $T$ the absolute temperature and
$V$ the volume of the sample. The computation of the effectiveness
starts in section S1 of the supporting information with the general
problem of the propagation of a distortion in junction fluctuations
in an otherwise perfect network. To this end, it is summed over the
modified elasticity of all attached strands in generation $g$ of
connected strands away from the loop, see Figure \ref{fig:Illustration-of-the}
for illustration. Thus, the result of section S1 of the supporting
information accounts for the impact of a loop on the surrounding network
structure. 

For computing the modified elastic effectiveness inside a loop made
of $i>1$ elastic strands, we reduce the loop inside the network to
the structure shown on the right of Figure \ref{fig:Illustration-of-the}:
Then, there is one strand $N_{\text{i}}$ that describes the connection
between the first and last cross-link of the cycle that runs not through
ground but through the rest of the finite cycle except of the $N$-mer,
while the elastic strands $K_{1}$ and $K_{\text{i}}$ describe the
effective connection to ground after removing all cross-links except
of cross-link 1 and $i$. The corresponding computations can be done
exactly using $Y-\Delta$ transforms \cite{Kenelly1899} and are described
in section S2 of the supporting information. 

With these results, the change in elastic effectiveness caused by
loops of arbitrary $i$ can be computed exactly within the ILGA. For
$i=1$, the total reduction in elastic effectiveness per pending loop
($i=1$) amounts to
\begin{equation}
\Delta\epsilon_{i=1}=1\label{eq:De1}
\end{equation}
independent of junction functionality $f$, see section S3 of the
supporting information. Equation (\ref{eq:De1}) agrees with the graph
theory approach of cutting dangling loops \cite{Flory1982} and similar
discussions in other works \cite{Dusek1978,Panukov1996}, but is in
contrast to ref. \cite{Zhong2016} that obtained a different result
for $f\le4$.

For $i=2$, it is found in section S4 of the supporting information,
that the net elastic effectiveness changes as 
\begin{equation}
\Delta\epsilon_{2}=\frac{2(f-1)}{f^{2}}.\label{eq:De2-1}
\end{equation}
This result agrees with ref. \cite{Zhong2016} after considering the
different normalization for elastic effectiveness and the coefficient
of $2/f$ for the difference in number density of network junctions
and strands for all $f$.

\begin{figure}
\includegraphics[angle=270,width=1\columnwidth]{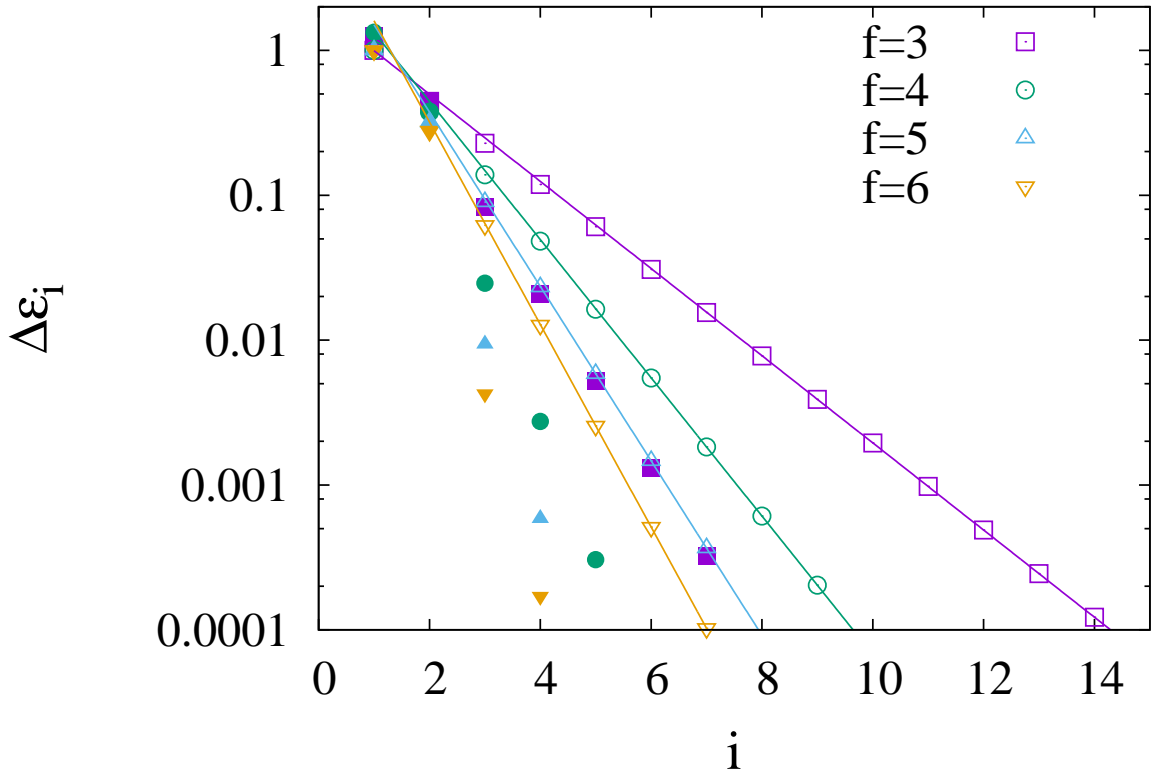}

\caption{\label{fig:Reduction-in-elastic}Reduction in elastic effectiveness
$\Delta\epsilon_{i}$ as a function of $i$. Open symbols is computation
of this work, full symbols is computation by ref. \cite{Zhong2016}.
The lines are a plot of equation (\ref{eq:Dei}).}
\end{figure}

Possible corrections due to finite loops were computed numerically
for $i>2$ and compared with the approximation of ref. \cite{Zhong2016}
in Figure \ref{fig:Reduction-in-elastic}. The exact solution approaches
for large $i$ quickly the approximation
\begin{equation}
\Delta\epsilon_{\text{i}}\approx\frac{2(f-2)}{(f-1)^{i}},\label{eq:Dei}
\end{equation}
that was included as lines in Figure \ref{fig:Reduction-in-elastic},
see section S4 of the supporting information for more details. The
exact solution of the present work is clearly larger than the approximate
result derived previously \cite{Zhong2016}, see Figure \ref{fig:Reduction-in-elastic},
but remains a small correction for large $i$. This difference compensates
to a large portion the over-estimate of the pending loops for $f\le4$
in previous work \cite{Zhong2016}. 

The above equations allow for a quantitative test of the impact of
finite loops on the elasticity of a network, if combined with statistical
information on the frequency of these finite loops \cite{Suematsu2002,Lang2005a,Wang2016}.
But as discussed below, finite loops are neither the only important
correction for the elasticity of model networks nor must the above
results provide a correct estimate for the corrections due to finite
loops in any situation. 

The weak point of the above results is that the derivation relies
solely on a consideration of cross-link fluctuations and how these
change when inserting a finite loop. As discussed in section S5 of
the supporting information, such an approach is fully legitimate when
linking instantaneously chains to a tree-like network when the ensemble
average chain conformations are in reference state. Then, the conformations
of the network chains can be split into a time average part and a
fluctuating part, whereby the latter is solely related to cross-link
fluctuations and the former to the time average stretch of the chains.
The portion by which this time average conformations are shorter than
the square reference chain size $R_{0}^{2}$ is identified as the
elastic effectiveness $\epsilon$ and determines the contribution
to modulus in multiples of $kT/V$.

However, such a line of arguments fails when considering finite loops.
The intricate point is that the ensemble average conformations of
a chain as part of a loop differ from the conformations of a linear
strand: the unperturbed average square size between two adjacent ``cross-links''
along an isolated loop of $i$ chains is equivalent to the unperturbed
size of a linear strand with 
\begin{equation}
N_{\text{a}}=N\frac{i-1}{i}\label{eq:N_a}
\end{equation}
segments. Thus, whenever a loop is built chain by chain as part of
a network (see section S6 of the supporting information for more details),
these strands enter the loop with unperturbed size but tend to contract
once the loop is closed to a square size that is smaller by a factor
of $(i-1)/i$. This creates extra stress on the surrounding network
chains that is not recognized when considering or measuring fluctuations,
since the fluctuations depend solely on connectivity (for a network
of Gaussian chains) and not on the spatial extension of the combined
strands. The contribution to modulus of a particular strand, however,
is proportional to the spatial extension of the time average strand,
see section S5 of the supporting information, and thus, proportional
to the extension of the corresponding combined strand for a given
$N$.

What complicates the discussion is that there is an alternative route
how a loop can become part of the network: if a loop is closed prior
to attaching it at least twice to the network, it can relax to the
average loop size. This latter route is important for small $i$ and
close to the gel point, while the former process dominates towards
the end of the reactions. Both ways of attaching strands lead to a
different contribution to elastic modulus, see section S6 of the supporting
information. But only the latter route agrees with the results that
were derived from considering cross-link fluctuations.

A second intricate point of randomly connected polymer networks is
that the chains inside a network are simultaneously part of several
loops, whereby the corresponding frequency distribution for participation
is rather broad, see Figure 5 of ref. \cite{Lang2003b} for an example.
Thus, the chains inside the network cannot accommodate all of the
constraints arising from the connected loops simultaneously. But one
could expect that the impact of the smallest loop where a chain is
part of stands out, since $N-N_{\text{a}}$ is largest for the smallest
loops. Thus, the number of strands $i_{\text{min}}$ of the smallest
loop where a network strand is part of needs to be determined. The
distribution of this $i_{\text{min}}$ is typically peaked at a small
$i_{\text{min}}$ in the range of 6-8 for end-linked model networks
with $f=4$ and an overlap number of $\approx8$ of an end-linked
network, see section S7 of the supporting information. Note that some
more information on loop size distributions for networks with different
structure and the corresponding peak position is available in ref.
\cite{Lang2007}.

\begin{figure}
\includegraphics[angle=270,width=1\columnwidth]{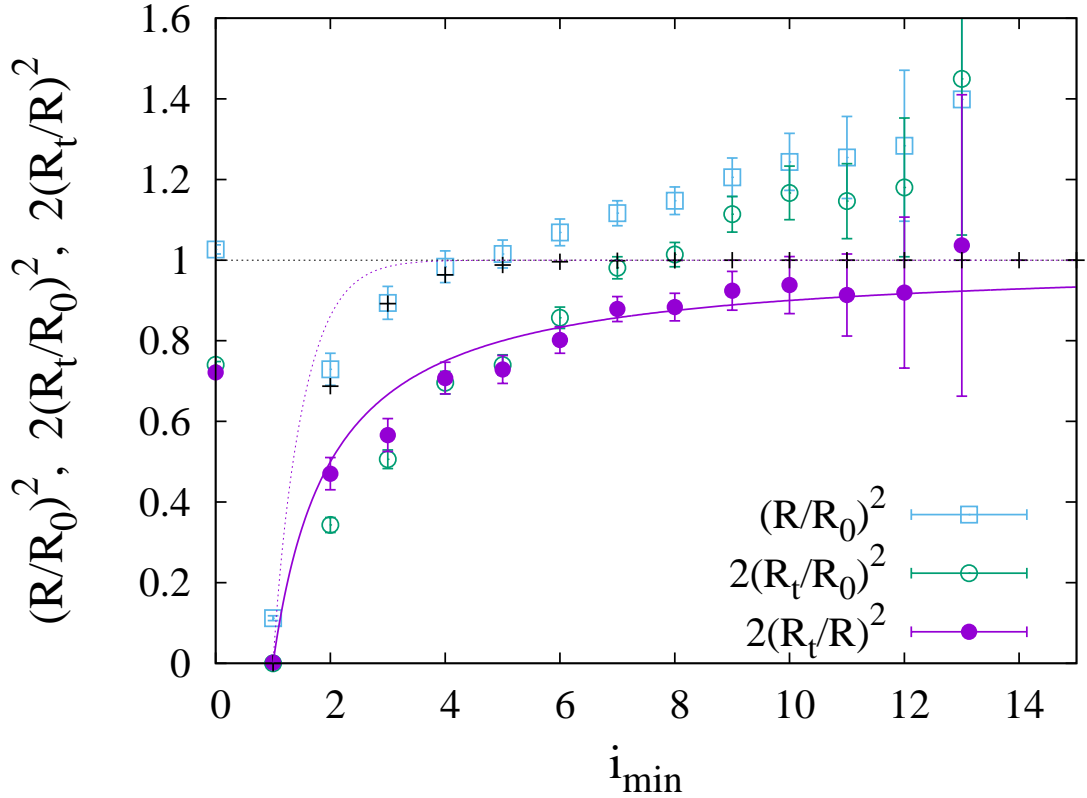}

\caption{\label{fig:Normalized-instantaneous-square}Normalized instantaneous
square size of active chains, $\left(R/R_{0}\right)^{2}$, twice the
normalized time average square size of active network chains $2\left(R_{t}/R_{0}\right)^{2}$,
and twice the ratio of time average and instantaneous size $2\left(R_{t}/R\right)^{2}$
as a function of $i_{min}$. The black crosses indicate the estimated
instantaneous chain size when considering the ILGA as discussed in
the supporting information. The averages over all $i_{min}$ are included
as extra data points at $i_{min}=0$. The continuous line is equation
(\ref{eq:N_a}), the red dashed line results from equation (11) of
ref. \cite{Zhong2016} that was considered as an approximation for
$i\ge3$.}
\end{figure}

The $i_{\text{min}}$ was determined for an untangled network where
the time average conformations of the chains have been analyzed previously
\cite{Lang2010,Lang2013}. Figure \ref{fig:Normalized-instantaneous-square}
compares the square ratios of instantaneous chain size, $R$, time
average chain size, $R_{\text{t}}$, and reference chain size in melt,
$R_{0}$. Averages over the whole network are included at $i_{\text{min}}=0$. 

The average chain size in the network appears to be almost not stretched
because of $\left(R/R_{0}\right)^{2}\approx1.03$, however, this observation
is partially biased by the contribution of pending loops. The fluctuating
chain size in non-pending loops $i>1$ is stretched by $\approx8\%$
as compared to the unperturbed size in melt. For this particular sample,
this correction is therefore larger than the impact of pending loops
that is here a correction of $\approx5\%$, see section S7 of the
supporting information. Now, let us ignore this stretching correction
and consider only the normalized time average size of all network
strands with $2\left(R_{t}/R_{0}\right)\approx0.74$. Consideration
of conversion of $p=0.95$ leads to an estimate of $2\left(R_{t}/R_{0}\right)\approx0.8$,
which is further reduced by the $5\%$ of pending loops to $2\left(R_{t}/R_{0}\right)\approx0.75$.
If compared with a measurement of modulus, one would probably conclude
that a consideration of conversion and pending loops is sufficient
to describe the data. But the stretched chain conformations and impact
of finite loops with $i>1$ would be missed by this simplified analysis.

A more detailed view is possible when looking into the dependence
of these ratios as a function of $i_{min}$ that indicate a striking
dependence on the loop size. Quite interestingly, when dividing out
stretch by normalizing time average size with respect to fluctuating
size, the corresponding ratio $2\left(R_{t}/R\right)$ approaches
roughly equation (\ref{eq:N_a}). This observation can be interpreted
such that all cyclic structures of the network attempt to maximize
entropy simultaneously by putting the surrounding chains and loops
under tension. Thus, indeed, one could consider an ``ideal gas of
loops'' as a reference state for the network. But this ideal gas
is under tension and one has to understand how this tension enters
the system or is being relaxed during the process of network formation
along with the development of the size distribution of loops as a
function of conversion.

In this context, it has to be mentioned that linking chains from non-reference
chain conformations has been discussed previously in the case of entangled
networks \cite{Lang2017}, but similar arguments apply for phantom
networks. The key idea is that towards full conversion, possible reaction
partners are predominantly found at a distance larger than reference
size, since otherwise, these reaction partners would have been encountered
previously by the compact exploration of space of reactive groups
attached to polymers \cite{DeGennes1982} for diffusion controlled
reactions. This leads to an extra stretch of the network strands as
a function of conversion. Note that also for reaction controlled systems,
a part of this bias towards more extended chain conformations must
survive. Thus, a thorough investigation of the phantom modulus of
random model networks needs to separate this conversion dependent
stretch from the cycle dependent part and should address the rate
dependence of reactions and relaxation processes. This is clearly
beyond the scope of the present letter that aims to point out the
main problems that have to be addressed.

Beyond the above problem of deformed chain conformations in a polymer
network, there are additional points that require more attention in
future work. First, these model networks are typically prepared by
a co-polymerization reaction. Composition fluctuations of the reacting
species freeze in beyond the gel point and may lead to a systematic
shift of the data away from a perfect mixing assumed in theory as
discussed previously \cite{Lang2012a}. Second, full conversion is
never reached but phantom modulus was not corrected for incomplete
conversion in ref. \cite{Zhong2016}. Note that earlier work on the
network disassembly spectrometry recognized a measurable amount of
pending chains \cite{Zhou2012} hinting in this direction. Finally,
the above results based upon the ILGA cannot rule out that there are
additional cooperative effects among connected loops, since real networks
are made entirely of finite loops. For these, we have no concepts
right at hand, but an analysis of the elastic properties of network
strands similar to Figure \ref{fig:Normalized-instantaneous-square}
might lead to an identification of such effects.

\section*{Acknowledgments}

The author acknowledges support from the DFG (grant LA2735/2-2) and
helpful discussions with A. Sharma, T. Müller, T. Kreer, K. Suresh
Kumar, and J.-U. Sommer.

\section*{Associated Content}

The supporting information contains a FORTRAN code in a separate file
that performs the computations as described in the supporting information.
The seven sections of the text document in the supporting information
concern: 1) The propagation of a network distortion and the corresponding
modification in elastic effectiveness of the surrounding network chains.
2) The reduction of non-pending loops of arbitrary number of chains
to loops containing two chains. 3) The elastic effectiveness of pending
cycles. 4) The elastic effectiveness of non-pending cycles as derived
from considering cross-link fluctuations only. 5) A simplified discussion
of the classical phantom model and how the cross-linking process is
taken into account. 6) The problem of how a loop is incorporated into
the network, which can cause elastic contributions different to the
ones estimated from cross-link fluctuations. 7) A test of principles
by comparing with preliminary simulation data.

\bibliographystyle{achemso}

\providecommand{\latin}[1]{#1}
\providecommand*\mcitethebibliography{\thebibliography}
\csname @ifundefined\endcsname{endmcitethebibliography}
  {\let\endmcitethebibliography\endthebibliography}{}

\newpage
\begin{figure}
\includegraphics[width=0.9\columnwidth]{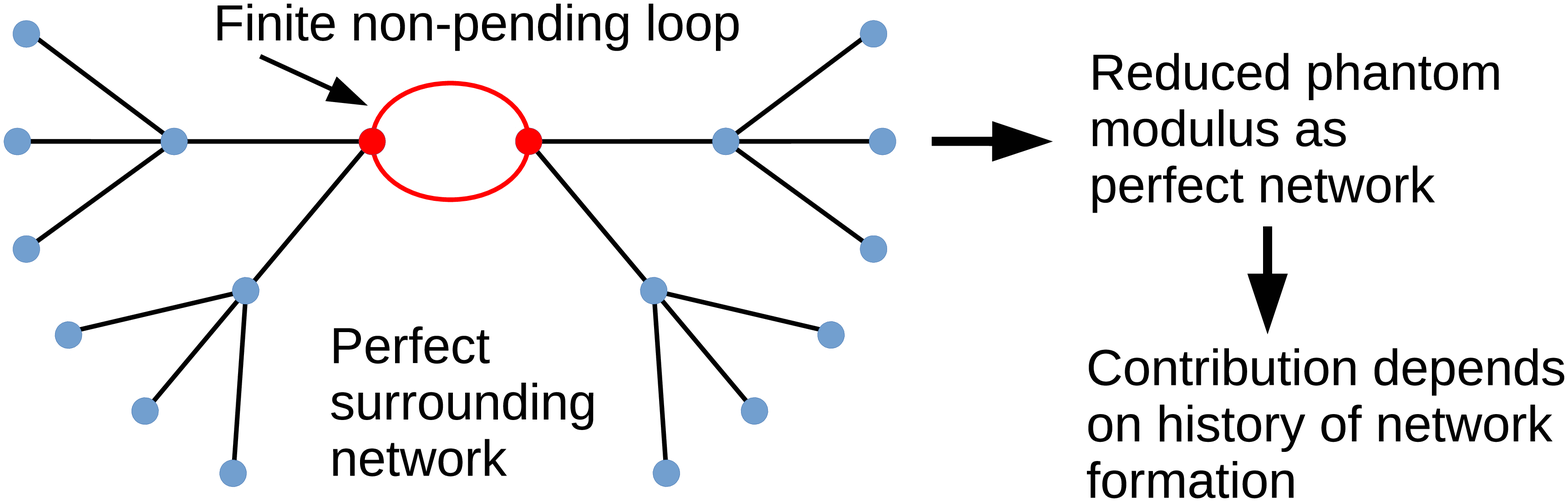}
\end{figure}

\end{document}